\documentclass[useAMS,usenatbib]{mn2e}
\usepackage{amsfonts}
\usepackage{amsmath}
\usepackage{amssymb}

\newcommand{\nn}{\nonumber}
\newcommand{\be}{\begin{equation}}
\newcommand{\ee}{\end{equation}}
\newcommand{\bea}{\begin{eqnarray}}
\newcommand{\eea}{\end{eqnarray}}
\newcommand{\baa}{\begin{align}}
\newcommand{\eaa}{\end{align}}
\newcommand{\rar}{\rightarrow}

\def\dd{ \text{d} }

\def\+{\dagger}
\def\la{\langle}
\def\ra{\rangle}
\def\parl{\parallel}
\def\n0{{\la n \ra}}
\def\dn{{\delta n}}
\def\B0{{\la B \ra}}
\def\dB{{\delta B}}

\newcommand{\cth}{\vartheta}

\def\ena{\end{eqnarray}}

\usepackage[dvips]{graphicx}
\usepackage{amsmath,amssymb}
\usepackage{myaasmacros}
\usepackage{color}
\usepackage{multirow}

\title{Mapping UHECRs deflections through the turbulent galactic magnetic field with the latest RM data}

\author[M. S. Pshirkov, P. G. Tinyakov, \& F. R. Urban]{M. S. Pshirkov$^{1,2,3}$\thanks{E-mail: pshirkov@prao.ru}, P. G. Tinyakov$^{3,4}$\thanks{E-mail: petr.tiniakov@ulb.ac.be}, F. R. Urban$^4$\thanks{E-mail: furban@ulb.ac.be}\\
$^{1}$Sternberg Astronomical Institute, 
Lomonosov Moscow State University, 
Universitetsky prospekt 13, 119992, Moscow, Russia\\
$^{2}$Pushchino Radio Astronomy Observatory, 142290 Pushchino, Russia\\
$^{3}$Institute for Nuclear Research of the Russian Academy of Sciences, 117312, Moscow, Russia\\
$^{4}$Universit\'e Libre de Bruxelles, Service de Physique Th\'eorique, CP225, 1050, Brussels, Belgium\\
}

\begin{document}

\date{}

\pubyear{2013}

\maketitle

\label{firstpage}

\begin{abstract}
We study the influence of the random part of the Galactic magnetic field on the propagation of ultra high-energy cosmic rays. Within very mild approximations about the properties of the electron density fluctuations in the Galaxy we are able to derive a clear and direct relation between the observed variance of rotation measures and the predicted cosmic ray deflections. Remarkably, this is obtained bypassing entirely the detailed knowledge of the magnetic properties of the turbulent plasma. Depending on the parameters of the electron density spectrum, we can either directly estimate the expected deflection, or constrain it from above. Thanks to the latest observational data on rotation measures, we build a direction-dependent map of such deflections. We find that over most of the sky the random deflections of 40 EeV protons do not exceed 1 to 2 degrees, and can be as large as 5 degrees close to the Galactic plane.
\end{abstract}


\section{Introduction}
\label{sec:intro}

Magnetic fields play a crucial role in the propagation of ultra-high energy
cosmic rays (UHECRs) --- cosmic rays with energies $E>10^{19}$ eV. If the
extragalactic magnetic fields are not too strong ($<1$ nG)~\citep{Kronberg:1993vk}, the influence of
the Galactic Magnetic Field (GMF) should be dominant. Various observations
indicate that the magnetic field of our Galaxy contains both regular and
turbulent components (see, e.g.,~\cite{Beck2001,Beck2008}). If there were only
a regular component, a fair knowledge of it would allow one to restore the
original positions of the sources from the observed arrival directions of
CRs. However, the presence of the turbulent part renders this task very
difficult. Unpredictable deflections of CRs in the turbulent fields sum up to
random displacements that smear the source image. Thus, the prospects to
reconstruct the actual positions of sources depend directly on the parameters
of the random part of GMF (hereafter, rGMF).

There is quite a substantial number of studies concerned with the question of
UHECR propagation through the GMF
(\cite{Stanev1997,Tinyakov2002,Prouza2003,Takami2008, Giacinti2010}, to name a
few). The influence of the random component is rather difficult to estimate
and the number of papers on this subject is
limited~\citep{Harari2002,Tinyakov2005,Golup2009}.

The rGMF affects not only the deflections of the CRs, but also contributes to
the scatter of the observed values of the Faraday Rotation Measures (RMs) of
extra-galactic radio-sources. Moreover, the rGMF enters both quantities in a
very similar fashion: they are proportional to the line-of-sight integrals of
the perpendicular components of rGMF (in the case of deflections) or the
parallel component times the electron density (in the case of RMs). Thus, with
some assumptions about the electron density, one can derive 
predictions/constraints on the random deflections directly from the observed
RM variances. The aim of this work is
to estimate the UHECR deflections due to rGMF using the recent observational
data on the RMs of extragalactic sources, namely the catalogue of RMs obtained
by~\citet{Taylor2009} from the NRAO VLA Sky Survey (NVSS), further simply
referred to as the NVSS RM catalogue.

The underlying logic is straightforward. In general, the variance of RMs over
a sufficiently small patch of the sky is composed of the part due to the
magnetic fields of our Galaxy (this is the part we are interested in) and
another part due to the intergalactic space, the RM intrinsic to the
sources, and due to the catalogue errors. The extragalactic contributions can be isolated and subtracted by
making use of their independence of the Galactic latitude. In turn, the
Galactic part of the RM variance is a sum of contributions due to rGMF and the
variation of the electron density. This relation allows one to either estimate
(if the second part can be neglected) or put an upper bound on the part due to
rGMF, which translates directly into an estimate/constraint on the UHECR
deflections. The result is a position-dependent map of
estimates/constraints on the UHECR random deflections. Such a map is
important in the UHECR data analysis, for instance, for a more accurate
prediction of UHECR flux from a given population of sources.

The paper is organised as follows. Section~\ref{sec:data} describes the
data. Section~\ref{sec:method} presents our method. The results and
conclusions are given in section~\ref{sec:results_and_conclusions}. The
closing Appendix contains all the technicalities of our method.

\section{Data}
\label{sec:data}

In this paper we used the NVSS RM catalogue consisting of 37,543 RM values of
extra-galactic sources. We cleaned the catalogue removing the outliers. The
algorithm for the cleaning procedure was the following: a circle of
$3^{\circ}$ radius was circumscribed around every source and the average RM
and its variance were calculated for the selected region. If the RM of the
source was more than two r.m.s.\ values away from the average value, the source
was marked as ``outlier''. In total, 1974 sources were removed after this
procedure, leaving 35,569 in the reliable set.

We binned the data using the HEALPix
package\footnote{http://healpix.jpl.nasa.gov }~\citep{healpix} into a map of
resolution $N_{\rm{side}}$ = 8 in the galactic coordinates with the ’RING’
pixel ordering. The total number of bins is equal to 768 and the area of each
bin is 53 sq.\ deg. This bin size provided the statistics of $\sim 50$ RMs in a
bin on average. At the same time, it is sufficiently small to neglect the
variation of the RM due to the change of the coherent magnetic field across
the bin.

Our resolution is thus that of our bins; if there is a very strong and localised regular magnetic field or strong variation of the regular electron density within such region, we would mistakenly interpret it as turbulent.  To refine our model one would need to remove the known strong features of the regular galactic field and electron density.  It would be also possible to exclude bins whose RM, compared with neighbouring bins, exceeds a given threshold.  However this introduces one more parameter in order to control the acceptance of the different bins.  We opt instead for a minor mistake in the interpretation of the result, but for a very simple logic, and also mathematical trackability.

\section{Method}
\label{sec:method}

\subsection{Instrinsic variance}
\label{sec:intrinsic-variance}

As already mentioned, the observed RM variance $\la RM^2 \ra \equiv \sigma$ consists of two main
parts\footnote{In this subsection \emph{only} we denote $\la RM^2 \ra \equiv \sigma$ so as to not clutter notation.}: the variance $\sigma_{\mathrm{ISM}}$ due to the interstellar medium
(ISM) we are interested in, and the variance $\sigma_{\mathrm{QSO}}$ that is
caused by other factors, including those intrinsic to the sources. We
estimated the variance $\sigma_{\mathrm{QSO}}$ using the data itself by making
use of the fact that the latter part should be independent of direction, while
the latitude ($b$) dependence of $\sigma_{\mathrm{ISM}}$ should look like $1/\sin{b}$ due 
to purely geometrical reasons. As we show below, a simple toy model of
identical cells of turbulent field up to some height $z_{\mathrm{thr}}$
predicts this dependence (see section~\ref{sec:toy-random-walk}).

Our method for separating the two variances had previously been introduced in~\citep{Schnitzeler2010}.  We have here slightly adapted this method for our purposes, and recalculated these variances anew: our results agree nicely with those obtained by Schnitzeler.

In order to estimate $\sigma_{\mathrm{QSO}}$ we performed the following
calculations. In the Northern hemisphere we fitted the observed $\sigma(b)$
for 36 uniformly spaced values $l=0^{\circ}, 10^{\circ}, ..., 350^{\circ}$
with the function $f(b)=({A^2/\sin^2{b}}+\sigma_{\mathrm{QSO}}^2)^{1/2}$
treating $A$ and $\sigma_{\mathrm{QSO}}$ as free parameters. The bins with
$b<10^{\circ}$ were excluded. We then calculated the mean and the r.m.s.\ values
of the resulting 36 values of $\sigma_{\mathrm{QSO}}$. 

The same procedure was independently performed for the Southern hemisphere. In
both cases we obtained very close numbers:
$\sigma_{\mathrm{QSO}}=(12\pm3)~\mathrm{rad~ m^{-2}}$. Thus, for the following
calculations we adopted $\sigma_{\mathrm{QSO}} = 12~ \mathrm {rad~ m^{-2}}$.
This value is largely dominated by the intrinsic errors of the NVSS RM catalogue $\sigma_{\mathrm{NVSS}}=10.4\pm0.4~ \mathrm {rad~ m^{-2}}$. 
 
Making use of this result, we estimated $\sigma_{\mathrm{ISM}}$ for each bin
as follows:
\begin{equation} \sigma_{\mathrm{ISM}} =
\begin{array}{r} \sqrt{\sigma^2-\sigma_{\mathrm{QSO}}^2},
~~\sigma>12 ~\mathrm{rad~m^{-2}} \end{array}. 
\label{sigma_ism}
\end{equation}
Both $\sigma$ and $\sigma_{\mathrm{QSO}}$ are determined with errors.  The
error of $\sigma_{\mathrm{QSO}}$ is $\delta \sigma_{\mathrm{QSO}} =3~ \mathrm
{rad~ m^{-2}}$ as explained above. The error of $\sigma$ depends on the number
$n$ of sources in the bin and equals $\delta \sigma=\sigma/\sqrt{2n})$. This
implies the error $\delta \sigma_{\mathrm{ISM}}$ in determination of
$\sigma_{\mathrm{ISM}}$,
\[
\delta\sigma_{\mathrm{ISM}}
=\sqrt{\delta\sigma_{\mathrm{QSO}}^2+\delta\sigma^2}.
\]
In those bins where the value of $\sigma_{\mathrm{ISM}}$ inferred from
eq.~(\ref{sigma_ism}) was smaller than $\delta\sigma_{\mathrm{ISM}}$ we have
set $\sigma_{\mathrm{ISM}} =\delta\sigma_{\mathrm{ISM}}$. This includes 29
bins with $\sigma<\sigma_{\mathrm{QSO}}$. 

Finally, we discarded the bins with the number of sources less than 10 (130 bins
total) located in the ``blind spot'' of the NVSS survey (declination
$<-40^{\circ}$) and its immediate vicinity.

\subsection{Rotation measures vs. deflections}
\label{sec:rotat-meas-defl}

When propagating through a magnetised plasma, the polarisation plane of a
linearly polarised electromagnetic wave of wavelength $\lambda$ rotates by the
angle $\Delta \psi$ proportional to the square of the wavelength,
\begin{equation}
\Delta\psi = RM \cdot \lambda^2, \label{RM1}
\end{equation}
where the coefficient RM, the rotation measure, can be expressed in terms of
the properties of the interstellar medium (ISM) and permeating magnetic
fields. Since the same magnetic field also deflects UHECRs, both
processes can be described by similar expressions, functions of the direction
$\hat r$:
\begin{align}
RM(\hat r) &= c_1 \int_0^D \dd z \, n \, B_\parl \, , \label{RM} \\
\cth_i(\hat r) &= c_2 \int_0^D \dd z \, \epsilon_{ij} \, B_j \, , 
\label{DA}
\end{align}
where $n$ is the electron density, $B_\parl$ and $B_i$, $i=(1,2)$ are the components of the magnetic field
parallel and perpendicular to the direction $\hat l$, respectively, $D$ is the
distance to the source, and the normalisation constants are $c_1 =
0.81\,\mathrm{rad}\,\mathrm{cm}^3/(\mathrm{m^2}\,\mathrm{pc}\,\mu\mathrm{G})
\approx 2.7\times10^{-23}\,\mathrm{rad}/\mu\mathrm{G}$,
$c_2=Ze/(E\mu\mathrm{G})$, $Z$ and $E$ being the UHECR charge and energy,
respectively. For our further estimates we assume that UHECRs are protons
($Z=1$) and have energy equal to 40 EeV, thence
$c_2=7.5\times10^{-24}~\mathrm{rad}/(\mu\mathrm{G}~\mathrm{cm})$. For further
reference, we also give the ratio
\begin{equation}
{c_2\over c_1} = 0.28 \, {\rm cm}^{-1}.
\label{eq:c2/c1}
\end{equation}
These estimates could be easily rescaled for any given composition and energy
of UHECRs.

The presence of the random component induces variances both on the observed
RMs and on the deflections of UHECRs. Although both mean deflections and mean RMs
are by definition zero for a random field, their r.m.s.\ are not.  The
variations in $RM$ are caused by fluctuations both in the magnetic field and the
electron density, whereas variations in $\cth$ are produced by magnetic
fields fluctuations alone. This makes the link between the two rather
contrived in general.

Let us separate the regular and random components as $n = \n0 + \dn$ and $B_i = \B0_i + \dB_i$, and similarly for $B_3 \equiv B_\parl$ --- all these quantities are taken to not depend on the absolute vertical distance to the observer, see the Appendix.  Conversely $RM$ and $\cth_i$ already refer to the (square roots of) their variances, and there will be no need to prepend a ``$\delta$'' to them.  With this notation, and within very mild approximations we can spell out the connection between $\la RM^2 \ra$ and $\la \cth^2 \ra \equiv \sum \la \cth_i \cth_i \ra$.

Schematically, for some positive coefficients $\alpha$ and $\beta$ of appropriate units, we show in the Appendix that we can write:
\begin{align}
\la RM^2 \ra &= c_1^2 \, \la n \ra^2 \la {\cal I}^2(\dB) \ra 
+ \alpha \, \la {\cal I}^2(\dB) \ra + \beta \, , \label{scheme}\\
\la \cth^2 \ra &= c_2^2 \, \la {\cal I}^2(\dB) \ra \, ,
\end{align}
cf.~(\ref{RM}) and~(\ref{DA}). We also have used the shorthand notation ${\cal I}(\star) \equiv \int \dd z \, \star$. In eq.~(\ref{scheme}), the first term comes from the variation $\dB$ alone, the second depends on both $\dB$ and $\delta n$ (the latter enters through the coefficient $\alpha$), while the third depends only on $\delta n$. The coefficients $\alpha$ and $\beta$ are both, therefore, proportional to $\la \delta n^2\ra$.

There are several parameters which enter the game at this stage, and impact
the values of the coefficients $\alpha$ and $\beta$. Before
dwelling upon these details, we see that there are two options. If the
properties of $\delta n$ are such that both $\alpha$ and $\beta$
are  small, the corresponding terms can be neglected. Then the magnetic
fluctuations factorise out altogether and we immediately infer
\begin{equation} 
\label{DefUL} 
\la \cth^2 \ra=\frac{c_2^2 }{c_1^2 \la n \ra^2 } \, \la RM^2 \ra.
\end{equation}
In this case we link directly (within the limits of the regular electron
density model) rotation measures with deflections.

In the opposite case of non-negligible $\alpha$ and $\beta$, there are additional, positive by construction, contributions to $\la RM^2 \ra$.  Then, the contribution of the magnetic field variance $\la \delta B^2 \ra$ alone to the RM one would be an overestimation, and~(\ref{DefUL}) remains valid although only as an upper limit.  We discuss the goodness of this upper bound below, but first, in Sec.~\ref{sec:toy-random-walk} we develop a simplified toy model which provides an additional intuitive argument to understand the physics behind~(\ref{DefUL}).

Before moving further, we need to mention a potential caveat of our approach.  Imagine that the regular magnetic field is strong enough as to deviate cosmic rays away from their paths by an angle which is significantly larger than the size of our bins.  If this is the case, then we are comparing two different lines of sight, the deflected cosmic ray, and polarised radio waves, which do not belong to the same bin.  We would be then inferring some properties of the former from the latter, but the two would not be causally connected, and the information extracted not meaningful.

Throughout our analysis we have in mind proton primaries, whose deflections at the energies of interest, as we checked, are significantly smaller than the size of our bins; hence, the complete algorithm works smoothly.  The limit of our simplification is of course in this assumption.

For heavier composition one needs to make sure that the UHECRs trajectories are not deflected too much.  Notice that in any case it would still be possible to apply our method, but first the UHECRs paths through the regular magnetic field need to be reconstructed, so as to assign each source to its appropriate bin.  This however introduces a significant complication and model-dependence (on the regular field).

One important reason we decided to focus on the simplest case for, is that all equations are derivable in a streamlined and logic way, where all the key parameters show up in the final expressions.  This makes it very versatile in understanding the physics, and in giving reasonable estimates for a wide range of parameters which to date are still very uncertain.  With the assumptions clearly spelt out, and the simple results at hand, it is easy to generalise to more complicated models once better data become available.

In this paper we have chosen to work with the model of~\cite{Pshirkov2011}.  We have explicitly checked that, for proton primaries, the perpendicular coherent deflections caused by the regular magnetic field do not cause the events to jump across to the wrong bin.

Given this key assumption we can perform our analysis in the simple way as we have outlined in this section.  It is important to notice that, \emph{a posteriori}, we can also check that the random deflections we finally obtain do not exceed the size of our bins, which makes our assumption self-consistent.

\subsection{A toy random walk model}
\label{sec:toy-random-walk}

The variance of the deflections experienced by UHECRs could be estimated
within a  simple cell model: the random field with given
strength $B_r$ is chopped into identical cells with size $l_\mathrm{cell}$; in each
cell the magnetic field is uniform and has some random orientation. The
electron density $n$ is assumed to be constant. RMs would then perform a
random walk along the line of sight, which we can put in formulas as
\begin{equation}
RM=k_1RM_0\sqrt{N},\label{sigma_rm}
\end{equation}
where $k_1$ is a numerical coefficient, $N$ is the total number of cells along
the line of sight, $RM_0$ is the average contribution of the elementary cell
--- one can think of this as~(\ref{RM}) with $\int \dd z \rar l_\mathrm{cell}$.

The deflections in this model are described by a very similar formula:
\begin{equation}
\cth=k_2\cth_0\sqrt{N},\label{sigma_def}
\end{equation}
where $k_2$ is another coefficient, and $\cth_0$ is the deflection 
in the elementary cell (once again, take~(\ref{DA}) and replace $\int \dd z \rar l_\mathrm{cell}$).

The exact value of the coefficients $k_1, k_2$ were calculated
numerically. First we simulated the random walk for the RM
accumulation. Taking into account that only the parallel component of the
magnetic field contributes proportionally to $\cos\phi_i$, where $\phi_i$ is
the angle between the random direction of the magnetic field and the direction
of the line of sight, then, $k_1$ reads
\begin{equation}
k_1=\frac{\left|\sum\limits_{i=1}^N \cos\phi_i \right|}{\sqrt{N}},\label{k1}
\end{equation} 
where $\phi_i$ is a random variable in $[0,\pi]$ range. The number of steps
$N$ was taken equal to 100,000 and we averaged over 10,000 realisations,
obtaining finally $$k_1=0.56.$$

The coefficient $k_2$ could be obtained in the very same fashion, but now
there is a two-dimensional random walk process and also it is the orthogonal
component of the magnetic field  that should be taken into account. One finds
\begin{equation}
k_2=\frac{\sqrt{\left(\sum\limits_{i=1}^N 
\sin \phi_i\cos \psi_i\right)^2
+\left(\sum\limits_{i=1}^N \sin\phi_i\sin\psi_i\right)^2}}{\sqrt{N}},
\label{k2}
\end{equation} 
where $\phi_i$ is again a random variable in the $[0,\pi]$ range, whereas
$\psi_i$ has domain $[0,2\pi]$.  We obtain
$$k_2=0.63.$$

Putting everything together, the relation between the RM and the deflection 
along the same path is
\begin{equation}
\cth \approx 0.31\,{\rm cm}^{-1}\frac{k_2}{k_1}\frac{RM}{\la n \ra} 
\approx 0.35\,{\rm cm}^{-1}\frac{RM}{\la n \ra} \, .
\end{equation}
This result is to be compared with the more accurate expression~(\ref{DefUL}) where, for $E=40$~EeV and $Z=1$ (protons), the coefficient turns out to be $c_2/c_1 \approx 0.28\,{\rm cm}^{-1}$, cf. eq.~(\ref{eq:c2/c1}). The two expressions are therefore in close agreement, the more so in light of the observational and theoretical errors we are faced with: the toy model correctly distillates the physics, and provides clear and intuitive insight for our idea.

\subsection{General case}
\label{sec:general-case}

We turn now to discussing the parameters which enter in $\alpha$ and $\beta$,
and the interpretation of~(\ref{DefUL}).  For concreteness we focus on the
simplest, and most corroborated by the data,
scenario~\citep{Armstrong:1995zc,Haverkorn:2008tb}.  We thus employ a
Kolmogorov electron density spectrum to obtain
\begin{align}
\label{RMcurve}
\la RM^2 \ra &\simeq c_1^2 \, \la n \ra^2 \la {\cal I}^2(\dB) \ra \\
&+ c_1^2 \left[3.1 C_n^2 l_0^{2/3} \la {\cal I}^2(\dB) \ra + 1.1 C_n^2 D l_0^{5/3} \la B \ra^2 \right] \, , \nn
\end{align}
where $C_n^2$ is the amplitude of the Kolmogorov spectrum, and the momentum
IR cutoff is given by $q_0 = 2\pi/l_0$, with $l_0$ the largest scale of
turbulence (with a given spectral index); finally, $D$ is the path length
along which the electron density turbulence extends (and, as we expect for the
magnetic field one, it will roughly follow a $1/\sin(b)$ distribution).  The
details of the calculation are given in the Appendix.

As previously mentioned, there are two possibilities.  If the magnetic field
variance dominates this expression, then there is a simple direct relation
$\la RM^2 \ra \propto \la {\cal I}(\dB)^2 \ra$.  The second option is that
the $\la B \ra^2$ term wins, in which case the measured RMs would be due to a
combination of fluctuations in the electron density and the magnetic field,
which we can not disentangle; the contribution of the magnetic field variance
$\la \delta B^2 \ra$ to the UHECR deflections would then be smaller than in
eq.~(\ref{DefUL}), so the latter would be an overestimation and would,
therefore, provide one with an upper bound on the deflections.

We can also turn this argument around.  Assume a given set of parameters
which describe the electron density, its fluctuations and the average magnetic
field.  We can ask which would be the contribution of the last term
(independent of the unknown $\la \delta B^2\ra$) to the measured RM. If this
is actually larger than what is observed, the conclusion is that the set of
parameters used is not consistent with the RM data. Still, eq.~(\ref{DefUL})
gives a valid upper bound for a given electron density.

Consider the parameters that enter eq.~(\ref{RMcurve}). First, we fix the value of the path length $D$ that should be inserted into our calculations. The variances get contributions only in regions where considerable magnetic fields and electron density are present. Both quantities are usually assumed to decay very rapidly with the height $h$ above the Galactic plane: like $\mathrm{sech}^2(h/d)$~\citep{Cordes2002}; or $\exp(-h/d)$~\citep{Gaensler2008}, albeit with possibly somewhat different vertical scales $d$. To describe the electron density distribution we use the NE2001 model~\citep{Cordes2002} modified with an increased vertical scale ($n(h=0)\approx0.014~\mathrm{cm^{-3}},~d = 1.8~\mathrm{kpc}$), which is implied by recent observations~\citep{Pshirkov2011} (see also~\citep{Gaensler2008}; in~\citep{Savage:2009hn,Taylor:1993my} instead lower values for $d$ were preferred: it is immediate to rescale all our results for any of these options, see below).  We adopt this vertical scale and cut the integrals at the value $D=d/\sin(b)$. Notice that since all three terms in eq.~(\ref{RMcurve}) are roughly proportional to $D$, their relative magnitude does not depend strongly on the adopted value.  Finally, we also assume that the vertical scale of the turbulent part of the GMF does not exceed $\sim 2$~kpc, i.e., magnetic field and electron density regions are largely spatially coincident (see Sec.~\ref{sec:results_and_conclusions} for more details).

Moreover, notice that we do not treat explicitly the $h$-dependence of the average $\la n \ra$ itself, which would also behave as $\mathrm{sech}^2(h/d)$ or $\exp(-h/d)$, although we do keep the dependence with direction.  In doing so we are making an approximation equivalent to substituting the steep suppression with a sharper step function localised at $h=d$, with the advantage that we can calculate everything analytically.  This boosts the deflection we obtain by at most around 30\% (this approximate value is obtained by comparing the integrals of a simple step function against the actual behaviour, however ignoring any $h/d$-dependence.), which is not large an error in view of the accuracies we have at hand.

Recall also our discussion at the end of Sec.~\ref{sec:data}: our results are interpreted as upper limits on the turbulent deflections.  If there are isolated, strong variations of the regular electron density field we would interpret them as coming from random fields: this does not impair the final result, and in principle such features can be removed to tighten the constraint we infer.

The mean magnetic field $\la B \ra$ entering eq.~(\ref{RMcurve}) can be
inferred from the existing models of the coherent Galactic field. We adopt
$\la B \ra$ estimated from the model~\citep{Pshirkov2011}; alternatively, one
could use, e.g., the model~\citep{Jansson:2012rt}.

The statistical properties of $\delta n$ are not very well known. The analyses of~\cite{Armstrong:1995zc} and~\cite{Haverkorn:2008tb} have obtained $C_n^2 = 10^{-3} m^{-20/3}$ and an IR scale of $l_0$ = 100 pc (at large $b$).  Again, these values are expected to change with direction because of the morphology of the galaxy, but the data are uncertain and we will rely on the simplifying approximation that these parameters remain the same across the sky.  These are our fiducial parameters, which we employ in drawing the deflection map.  With these values, for a typical average magnetic field of 1~$\mu$G and $b=90^\circ$, the third term in eq.~(\ref{RMcurve}) amounts to about 17 rad/m$^2$, which overshoots the observed RM in a large part of the sky, although not by much. We have therefore a situation described above when the parameters are not quite consistent with the observed RMs. If this inconsistency is attributed to the numerical uncertainties, one concludes that the third term in eq.~(\ref{RMcurve}) dominates and thus eq.~(\ref{DefUL}) is an overestimate.

We have also tried different parameters set, besides the blueprint values mentioned above.  The first possible modification is to rescale the vertical scale of electron density to 1.3 kpc instead of 1.8 kpc, as suggested by~\cite{Schnitzeler:2012jq}.  The main effect in this case is that $\la n \ra$ is somewhat increased, which implies that the resulting deflections will be smaller.  The physics does not change hoever, and the same qualitative conclusions can be drawn.  The displacement angles are diminished by at most around 20\% in some parts of the sky, which is not significant within our approximations.

Then, so far we have always assumed exact Kolmogorov turbulence all the way up to the IR cutoff $l_0$ of 100 pc.  Nevertheless, the observational scenario is far from being clear, and there is the possibility that the spectrum actually flattens at larger scales; there are indications of this flattening for the \emph{turbulent magnetic field}~\citep{Regis:2011ji}, although it is uncertain what the link between the two is at large scales.  Within our formalism it is immediate to estimate its effects: as a concrete case, taking the spectral index $\alpha = 3$ from 1 pc to 100 pc suppresses the contribution of the third term in eq.~(\ref{RMcurve}) to the total RMs by a factor of $100^{-1/3}$, bringing it down to about 4 rad/m$^2$, which is below the intrinsic threshold we adopt.  In this case~(\ref{DefUL}) becomes a much better estimation of the deflections: if we ask that the third term in eq.~(\ref{RMcurve}) contributes at most to 20\% of the observed RM, then any RM above about 20 rad/m$^2$ would provide a direct connection to $\la\cth^2\ra$ (within our approximations). Let us emphasise however that we have stuck to the baseline parameters for the map, and ensuing discussion.

The last comment is on the second term in eq.~(\ref{RMcurve}). This term has the same dependence on $\la\delta B^2\ra$ as the first one, so the two can in principle be combined together, which would change (make smaller) the coefficient in eq.~(\ref{DefUL}). Although this would strengthen the constraints on the deflections for a given value of RM, it would also make them dependent on the (poorly known) properties of the electron density fluctuations. For this reason we do not include this term in our final constraints. Using again the fiducial values of~\cite{Armstrong:1995zc} and~\cite{Haverkorn:2008tb} we find $1.76 C_n l_0^{1/3} \sim 0.080\mathrm{cm}^{-3}$, which is larger than the model average of NE2001; the latter is,  $\la n\ra \approx 0.012\,\mathrm{cm}^{-3}$).  Using instead the spectral index $\alpha = 3$ from 1 pc to 100 pc suppresses this contribution down to $0.017\,\mathrm{cm}^{-3}$, thereby bringing it around the mean value of NE2001.

To summarise, within present uncertainties of the parameters of the electron
density fluctuations, both third and second terms in eq.~(\ref{RMcurve}) can
be sizeable, but their values cannot be reliably determined at present. 
Because of these uncertainties, we base our
constraints on eq.~(\ref{DefUL}) which we treat as an upper bound on
deflections $\la \cth^2 \ra$.

\section{Results and conclusions}
\label{sec:results_and_conclusions}

As explained above, eq.~(\ref{DefUL}) provides one with an upper bound (or an
estimate, if the contributions of the fluctuations of the electron density can
be neglected) on the UHECR deflections $\la \cth^2 \ra$ in the random
component of the Galactic magnetic field. These deflections can be translated
into the experimentally observable displacements $\la \Theta^2 \ra$ of the
UHECR arrival directions at the Earth with respect to the direction to the
true position of the source. The relation between the two quantities
is~\citep{Harari2002}:
\begin{equation}
\la\Theta^2\ra = \la \cth^2 \ra / 3 \, .
\end{equation}

\begin{figure}
\begin{center}
\includegraphics[width=8 cm]{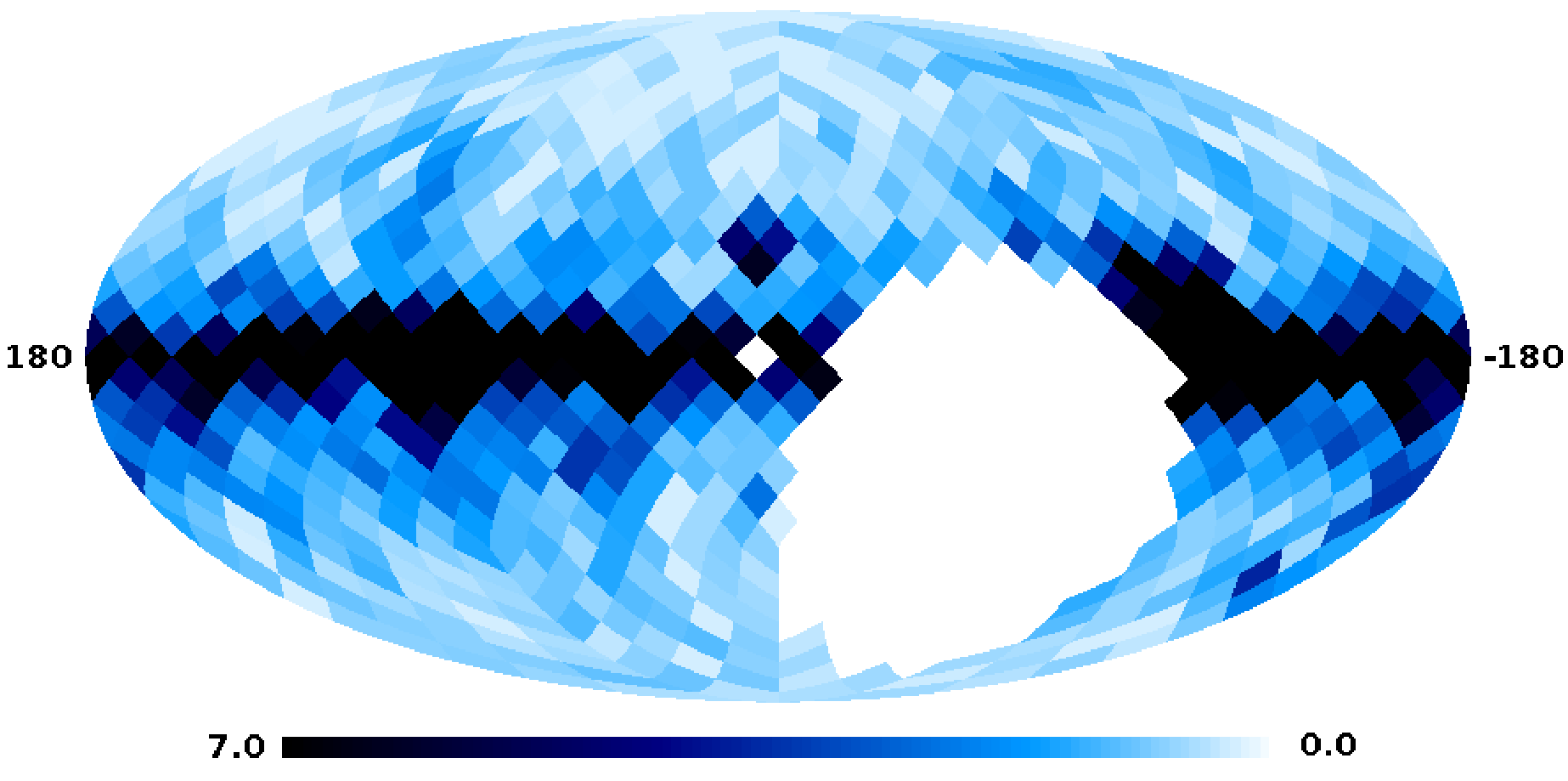}\\
\end{center}
\caption{The map of displacements $\sqrt{\la \Theta^2 \ra}$, of protons with
  energy $E=4\times10^{19}$ eV due to the random component of the GMF as
  follows from eq.~(\ref{DefUL}).  The map is in Galactic coordinates and the
  displacements are in degrees.  This map should be considered as an upper
  bound and may turn to an estimate if the contribution of the electron
  density fluctuations are small, as explained in the text.
\label{fig:map}}
\end{figure}

In Fig.~\ref{fig:map} we show the map of deflections $\sqrt{\la \Theta^2 \ra}$
of protons with energy $E=4\times10^{19}$ eV due to the random component of
the GMF calculated according to eq.~(\ref{DefUL}) where we have used the value
$c_2/c_1=0.28\,{\rm cm}^{-1}$. This map should be considered an upper
bound, which may turn into an estimate if the contribution of the electron
density fluctuations will be shown to be small in the future.
The displacements $\sqrt{\la \Theta^2 \ra}$ are bound to 1 to 2 degrees
for the majority of directions, although closer to the galactic plane the
bounds are relaxed to $\sim 5^\circ$. This map is the main result of our
paper. 

One can see from Fig.~\ref{fig:map} that the limit on deflections does not
depend on the Galactic longitude in any regular way. In particular, the
direction towards the Galactic centre does not look any special. On the
contrary, there is a regular dependence on the Galactic latitude $b$. This
dependence is fitted well with the phenomenological function $A/(\sin^2(b)+B)$, with the
constants $A=0.5$, $B=0.13$, although the fluctuations are large. For the
purpose of describing the upper bound, the constraints of Fig.~\ref{fig:map} 
can be majorated by the analytic function 
\begin{equation}
\sqrt{\la \Theta^2 \ra} < {1^\circ \over \sin^2(b) + 0.15} 
\label{eq:majoration}
\end{equation}
everywhere except several bins. 

The results obtained in the paper are in good agreement with results of~\cite{Tinyakov2005}.  In the latter paper the deflections in the turbulent field were derived using 
properties of the rGMF inferred from few observations coming from different sets of data.  It was shown that for 40 EeV protons such deflections are small, not exceeding $1.5^{\circ}$ for the better part of the sky.  The most important difference with that work, where one was obliged to use a specific model (i.e., spectrum) for the rGMF fluctuations to draw some quantitative conclusion, is that here we were able to place robust upper limits from a single, thereby consistent, set of data.

A word of warning is in order. While the constraints shown in
Fig.~\ref{fig:map} are conservative in the sense that we have treated all the
uncertainties in a way that increases the deflections (and thus weakens the
constraints), they only apply to the contribution of the region of $\pm 2$~kpc
around the Galactic disk where the electron density is large and constrained
by observations. One cannot exclude, by the arguments presented in the paper,
that there might exist more extended regions around the Galaxy that carry
small electron density and substantial magnetic fields such that their
contributions to the RMs are small, but the contribution to deflections of
UHECRs is larger than in eq.~(\ref{DefUL}) like in the model
of~\cite{Ahn:1999jd}.  Constraints on such models require a different approach,
which we leave for future study.

One needs to bear in mind that one technical limitation of our method is related to our resolution; for instance, the dark spot just to the left and north of the Galactic Centre is caused by the HII region surrounding the star Zeta Ophiuchi~\citep{HarveySmith:2011fe}, and is a large-scale field that shows structure on scales smaller than the cell size we use.  This would na\"ively incorrectly be counted as a particularly strong random fluctuation.  To refine the map all such known strong features should be removed.  However this introduces a high degree of uncertainty and does not add much to the main conclusions; the interpretation of the map as an upper bound is anyhow not compromised.

To conclude, there are three main lessons which we have learnt from this analysis.  Firstly, the deflections across the sky do not exhibit strong feature in Galactic longitude; it is not possible to tell the Galactic centre apart from the anticentre, for example.  Secondly, there is instead a clear regularity in Galactic longitude; we have provided a simple phenomenological fit which is of direct implementation for UHECRs anisotropy studies.  Thirdly, the overall magnitude, for high energy proton primaries of $E\gtrsim40$ EeV, does not exceed $5^\circ$; in fact, it is typically quite smaller over most of the sky --- especially away from the plane.

The use of RM data to constrain UHECRs deflections has thus revealed to be intuitive, solid, efficient, and effective, within the approximations of the method; a method which is easily and readily generalised and improved, accounting for any new data for modelling electron density and galactic magnetic field.

\section*{Acknowledgements}

This research has made use of NASA's Astrophysics Data System. Some of the results in this paper have been derived using the HEALPix (K.~M.~G\'orski et al., 2005, ApJ, 622, p759) package. The authors thank M.~Haverkorn for useful correspondence, S.~Troitsky for comments on the manuscript, and D.~Schnitzeler for constructive remarks. FU wishes to thank the Institute for Theoretical Astrophysics of Oslo, where part of this work was completed. The work of MP is supported by RFBR Grants No.~12-02-31776 mol\_a, No.~13-02-00184a, No.~13-02-01311a, No.~13-02-01293a, by the Grant of the President of Russian Federation MK-2138.2013.2 and by the Dynasty Foundation. FU and PT are supported by IISN project No.~4.4502.13 and Belgian Science Policy under IAP VII/37.

\section*{Appendix}
\label{appendix}

We describe here in details the formal steps that read to the relation~(\ref{DefUL}) between the variances of RM and deflections.

\subsection*{Definitions}
The two quantities we want to calculate are
\begin{flalign*}
& RM(\hat r) = c_1 \int_0^D \dd z \, n \, B_\parl \, , &\\
& \cth_i(\hat r) = c_2 \int_0^D \dd z \, \epsilon_{ij} \, B_j \, ; &
\end{flalign*}
more specifically, we need the variances of these quantities.  In principle, both the random variations in electron density and magnetic field are present.  Thus, we split $n = \n0 + \dn$ and $B_i = \B0_i + \dB_i$, where $i=(1,2)$ denotes the two components perpendicular to the line of sight, and similarly for $B_3 \equiv B_\parl$.

In what follows we approximate the actual distance (from the observer) dependence of such quantities is that of a pure step function.  This is certainly a good approximation as it captures its key feature: a sharp drop beyond a given distance $D$.  Indeed, we do expect that both the electron density and magnetic field (regular components and fluctuations alike) follow the smoothed distribution of ionised gas in the galaxy, which is typically coded as $\mathrm{sech}^2(x)$~\citep{Taylor:1993my,Cordes2002} or $\exp(-x)$~\citep{1985MNRAS.213..613L,Gaensler2008} ($x$ is the vertical distance from the observer normalised to $d$); both these functions do not vary exageratedly up to $x\sim1$ and rapidly decay beyond that point --- the mistake we are making in treating these as step functions is of, at most, around 30\%, which is well below our other uncertainties.  We are therefore allowed to average, along the line of sight, the regular components $\la n \ra$ and $\la B \ra$ within the integrals and bring them outside.  This also implies that the spectra of fluctuations will not depend on the absolute position, but only on the scale.

Thus, we define
\begin{flalign*}
I_{ij}	& = \la \dB_i(\bar{r}_0)\dB_j(\bar{r}_0 + \bar{r}) \ra &\\
		& = \int \dd^3q e^{-i\bar{r}\bar{q}} \frac{P_B(q)}{2q^3} \left( \cth_{ij}-\frac{q_i q_j}{q^2} \right) \, , &
\end{flalign*}
where we used the Fourier transform
\begin{flalign*}
& \dB_i(\bar{r}) = \int \dd^3q e^{i\bar{r}\bar{q}} \dB_i(\bar{q}) \, , &
\end{flalign*}
and the spectrum is defined as
\begin{flalign*}
& \la \dB_i(\bar{q})\dB_j^*(\bar{p}) \ra = \frac{P_B(q)}{2q^3} \left( \cth_{ij}-\frac{q_i q_j}{q^2} \right) \delta^3(\bar{q}-\bar{p}) \, . &
\end{flalign*}
Similar definitions apply to $\dn$:
\begin{flalign*}
& \dn(\bar{r}) = \int \dd^3q e^{i\bar{r}\bar{q}} \dn(\bar{q}) \, , &
\end{flalign*}
and
\begin{flalign*}
& \la \dn(\bar{q})\dn^*(\bar{p}) \ra = \frac{P_n(q)}{q^3} \delta^3(\bar{q}-\bar{p}) \, . &
\end{flalign*}
Notice that $I_{ij} \propto \cth_{ij}$ since otherwise it is zero (parity).  Writing the integrals explicitly, they all reduce to the prototype
\begin{flalign*}
& I_{i} = \int \dd\cos\cth \dd\cth \dd q q^2 e^{-irq\cos\cth} \frac{P_B(q)}{2q^3} \left( 1-\frac{q_i^2}{q^2} \right) \, , &
\end{flalign*}
where we call $z$ the line of sight axis.  The angular integrals can be solved to give
\begin{flalign*}
& I_1 = I_2 = 2\pi \int \frac{\dd q}{q} P_B(q) \left( \frac{\sin qr}{qr} + \frac{\cos qr}{q^2r^2} - \frac{\sin qr}{q^3r^3} \right) \, , &\\
& I_3 = I_\parl = 4\pi \int \frac{\dd q}{q} P_B(q) \left( \frac{\cos qr}{q^2r^2} - \frac{\sin qr}{q^3r^3} \right) \, . &
\end{flalign*}
Our assumptions on the properties of these fluctuations with vertical distance amount to assuming that the spectra we have hitherto defined do not depend on such quantity, but only on the (modulus of the) correlation length $r$.

The next step is to integrate along the line of sight:
\begin{flalign*}
\la \cth_i\cth_i \ra	& = c_2^2 \int_0^D \dd z \dd z' \epsilon_{ij} \epsilon_{ik} \la \dB_j(z)\dB_k(z') \ra &\\
							& = 2c_2^2 \int_0^D \dd u \int_0^u \dd r (I_1+I_2) &\\
							& = 4\pi c_2^2 \int \frac{\dd q}{q} P_B(q) \left( qD Si(qD) + \cos qD - \frac{\sin qD}{qD} \right) \, ; &
\end{flalign*}
here we have defined $r=z'-z$ and $u=z'+z$, and $Si(x)$ is the sine-integral function.

A very similar result can be obtained for the RM.  The full expression for the variance of $RM$, containing all pieces, reads
\begin{flalign}\label{almighty}
\la RM^2 \ra	& = c_1^2 \int_0^D \dd z \dd z' \bigg[ \n0^2 \la \dB_\parl(z)\dB_\parl(z') \ra &\\
				& + \la \dn(z)\dn(z') \ra \B0_\parl^2 + \la (\dB_\parl\dn)(z)(\dB_\parl\dn)(z') \ra \bigg] \, . \nn&
\end{flalign}
where the first term, $\la \dB_\parl^2 \ra$, is the contribution of rGMF proper, the second one, $\la \delta n^2 \ra$, is that of the electron density fluctuations alone, and the last one represents the convolved variations.

Notice that in this scheme we need to compare two quantities whose lines of sight are ``compatible'', which in our language means they must reside in the same bin.  If the regular magnetic field is strong enough as to deviate cosmic rays away from their paths by an angle which is significantly larger than the size of our bins, then the two different trajectories, the deflected cosmic ray, and polarised radio waves, are not causally connected, and our method does not apply.  There is no conceptual obstacle to taking this effect fully into account, by mapping the deflected UHECRs paths on the sphere through the regular field; these reconstructed lines of sight can then be employed for the analysis.  We do not do so, for we work with proton primaries whose deflections are small.  More details are given in the main text.

\subsection*{A simplified result}
If the variance on the electron density fluctuations is small, then the only significant term left in the lengthy expression~(\ref{almighty}) is the first, leading to
\begin{flalign*}
& \la RM^2 \ra = c_1^2 \n0^2 \int_0^D \dd z \dd z' \la \dB_\parl(z)\dB_\parl(z') \ra &\\
& \,\,\,\, = 4\pi c_1^2 \int \frac{\dd q}{q} P_B(q) \left( qD Si(qD) + \cos qD - 2 + \frac{\sin qD}{qD} \right) \, . & 
\end{flalign*}
To further simplify this expression, one can expand for large distances and small coherence length $qD\gg1$ to obtain
\begin{flalign*}
&\la \cth^2 \ra = 4\pi c_2^2 \int \frac{\dd q}{q^3} P_B(q) \left( \frac{\pi}{2} qD + \ldots \right) \, , &\\
&\la RM^2 \ra = 4\pi c_1^2 \n0^2 \int \frac{\dd q}{q^3} P_B(q) \left( \frac{\pi}{2} qD + \ldots \right) \, . &
\end{flalign*}
This demonstrates now the two quantities are, in first approximation, directly related, and one can infer the deflections caused to the UHECRs \emph{without knowing anything about the spectrum of the magnetic field}.  Of course this result makes use of the assumption that there are not any large electron density fluctuations, the goodness of which we discuss in the next section.

\subsection*{The convolution}
We now want to determine the contribution of the convolution
\begin{flalign*}
& \la (\dB_\parl\dn)(z)(\dB_\parl\dn)(z') \ra &\\
& \quad = \int \dd^3k \dd^3k' e^{-i\bar{k}\bar{z}-i\bar{k}'\bar{z}'} \la (\dB_\parl\dn)(\bar{k})(\dB_\parl\dn)(\bar{k}') \ra &\\
& \quad = \int \dd^3p \dd^3q e^{-i(\bar{q}+\bar{p})(\bar{z}'-\bar{z})} \frac{P_n(p)}{p^3} \frac{P_B(q)}{2q^3} \left( \cth_\parl - \frac{q_\parl^2}{q^2} \right) &\\
& \quad = (4\pi)^2 \int \frac{\dd p}{p} \frac{\dd q}{q} P_n(p) P_B(q) \frac{\sin pr}{pr} \left( \frac{\sin qr}{q^3r^3} - \frac{\cos qr}{q^2r^2} \right) \, . &
\end{flalign*}
This has to be integrated twice over $z$ and $z'$, or $r$ and $u$ as usual.  One reasonable assumption, which simplifies greatly the (euphemistically) unwieldy result, is that the integral is going to be dominated by the IR cutoffs, since observations tell us that the spectra are almost certainly red.  The integral can then be split in two, for $p \ll q$ and $p \gg q$ depending on which IR cutoff is the most important one.  The result can (now) be expanded for large $qD$ if this is the dominant cutoff, or, conversely for large $pD$ to find
\begin{flalign*}
& \la RM^2 \ra \subset (2\pi)^3 c_1^2 \int \frac{\dd p}{p} \frac{\dd q}{q} P_n(p) P_B(q)
	\begin{cases}
	\frac{2}{3} \frac{D}{q} & p \ll q \, , \\
	\frac{D}{p} & p \gg q \, .
	\end{cases} &
\end{flalign*}

Compare these with the integrated $\la \dn^2 \ra$ and $\la \dB^2 \ra$ in the same limit of large $pD$ and $qD$ --- recall that ${\cal I}(\star) \equiv \int \dd z \, \star$:
\begin{flalign*}
& \la {\cal I}(\dn)^2 \ra	= -4\pi^2 \int \frac{\dd p}{p} P_n(p) \frac{D}{p} \, , &\\
& \la {\cal I}(\dB)^2	\ra = -2\pi^2 \int \frac{\dd q}{q} P_B(q) \frac{D}{q} \, : &
\end{flalign*}
one can write the convolution part as either something proportional to one or the other.  In any case, the two limits differ only by a factor of 2/3, so we can take either one: the mistake we make is at most this much.  Physically, the first option might be more justified, that is, turbulence in electron density extends beyond, at larger scales, that for the magnetic field, so, for concreteness, we choose this in what follows.  The main result then reads
\begin{flalign*}
& \la {\cal I}(\dn\dB)^2 \ra = -\frac{8\pi}{3} \la {\cal I}(\dB)^2 \ra \int \frac{\dd p}{p} P_n(p) \, . &
\end{flalign*}
As this relation illustrates, everything can be predicted with the mild
assumption that we know the features of the electron density spectrum, and
there is no necessity to specify the properties of the magnetic field random
fluctuations.  Notice that thus far the only approximation we have employed is
that the IR dominates the fluctuation spectra.

\subsection*{Results}
We need to know the average electron density $\la n \ra$ and regular magnetic
field $\la B \ra$, and the spectrum of the electron density fluctuations
$P_n(q)$, in order to be able to extract the deflections from the RM data
bypassing the magnetic field fluctuations.

The spectrum $P_n$ can be written as
\begin{flalign*}
& \la \dn^2 \ra(\bar{r}) = \int \dd^3 q \frac{C_n^2 e^{-i\bar{r}\bar{q}}}{(q_0^2+q^2)^{\alpha/2}} \, , &
\end{flalign*}
where $C_n^2$ is the amplitude of a Kolmogorov spectrum for the electron density fluctuations, and the momentum IR cutoff is given by $q_0 = 2\pi/l_0$, with $l_0$ the largest scale of turbulence (with a given spectral index).  This integral can be solved exactly, and it gives
\begin{flalign*}
& \la \dn^2 \ra(r) = \frac{2\pi^{3/2}C_n^2}{\Gamma(\alpha/2)} \left( \frac{r}{2q_0} \right)^{(\alpha-3)/2} K_{(\alpha-3)/2}(q_0r) \, . &
\end{flalign*}
These definitions enable us to compute all the terms in~(\ref{almighty}).  As an example, following the results of~\cite{Armstrong:1995zc} and~\cite{Haverkorn:2008tb}, for Kolmogorov electron density fluctuations ($\alpha = 11/3$), we find
\begin{flalign*}
& \la RM^2 \ra \simeq &\\
& \qquad c_1^2 \left\{ \left[ \la n \ra^2 + 3.1 C_n^2 l_0^{2/3} \right] \la {\cal I}^2(\dB) \ra + 1.1 C_n^2 D l_0^{5/3} \la B \ra^2 \right\} \, , &
\end{flalign*}
which is the result~(\ref{RMcurve}) reported in the main text, and justifies the use of~(\ref{scheme}).  The ensuing discussion then follows.

\bibliographystyle{mn2e}
\bibliography{Deflection}
\end{document}